\documentclass[useAMS,usenatbib]{mn2e}

\usepackage{amsmath}
\usepackage{amssymb}
\usepackage[dvipsnames]{xcolor}
\usepackage{graphicx}
\usepackage{graphics}
\usepackage{multirow}
\usepackage{multicol}
\usepackage{dirtytalk}
\usepackage{hyperref}
\hypersetup{
     colorlinks   = true,
     citecolor    = blue
}

\def\apj{{ ApJ}}
\def\apjl{{ApJL}}
\def\apjs{{ ApJS}}

\def\aap{{ A\&A}}
\def\aj{{ AJ}}
\def\mnras{{ MNRAS}}
\def\araa{{ ARA\&A}}

\def\nat {{ Nature}}

\def\ssr{{ Space Sci. Rev.}}

\def\physrep{{ Physics Reports}}

\def\prd{{ Phys. Rev. D}}

\def\mr{\mathrm}

\def\mc{\mathcal}
\def\d{\mr{d}}
\def\b{\mathbf}
\def\me{m_{\rm e}}
\def\t{\widetilde}
\def\eps{\mc{E}}

\def\msun{M_{\rm \odot}}
\def\rsun{R_{\rm \odot}}
\def\ve{v_{\rm e}}
\def\ab{a_{\rm b}}
\def\rp{r_{\rm p}}

\def\rh{r_{\rm h}}
\def\rT{r_{\rm T}}
\def\rTe{r_{\rm T,e}}
\def\rTb{r_{\rm T,b}}
\def\mb{m_{\rm b}}
\def\me{m_{\rm e}}
\def\Rb{R_{\rm b}}

\newcommand{\lara}[1]{{\langle#1\rangle}}

\title[Hyper-Velocity Star]{The former companion of the hyper-velocity star S5-HVS1}
\author[Lu et al.]{Wenbin
  Lu$^{1}$\thanks{wenbinlu@caltech.edu},
  Jim Fuller$^{1}$,
  Yael Raveh$^{2}$,
  Hagai B. Perets$^{2}$,
  Ting S. Li$^{3,4}$\thanks{NHFP Einstein Fellow},
  \newauthor
  Matthew W. Hosek Jr.$^5$, and Tuan Do$^5$
\\
$^{1}$Theoretical Astrophysics, Walter Burke Institute for Theoretical Physics, Mail Code
350-17, Caltech, Pasadena, CA 91125, USA \\
$^{2}$ Physics Department, Technion --- Israel Institute of Technology, Haifa 3200003, Israel\\
$^{3}$ Observatories of the Carnegie Institution for Science, 813 Santa Barbara St., Pasadena, CA 91101, USA\\
$^{4}$ Department of Astrophysical Sciences, Princeton University, Princeton, NJ 08544, USA\\
$^{5}$ UCLA Galactic Center Group, Physics and Astronomy Department, University of California, Los Angeles, CA 90024, USA
}

\begin{document}
\label{firstpage}
\maketitle

\begin{abstract}
The hyper-velocity star S5-HVS1, ejected 5 Myr ago from the Galactic Center at 1800 km/s, was most likely produced by tidal break-up of a tight binary by the supermassive black hole SgrA*. Taking a Monte Carlo approach, we show that the former companion of S5-HVS1 was likely a main-sequence star between 1.2 and $6\msun$ and was captured into a highly eccentric orbit with pericenter distance in the range 1--10~AU and semimajor axis about $10^3$~AU. We then explore the fate of the captured star. We find that the heat deposited by tidally excited stellar oscillation modes leads to runaway disruption if the pericenter distance is smaller than about $3\rm\, AU$. Over the past 5 Myr, its angular momentum has been significantly modified by orbital relaxation, which may stochastically drive the pericenter inwards below $3\rm\, AU$ and cause tidal disruption. We find an overall survival probability in the range  5\% to 50\%, depending on the local relaxation time in the close environment of the captured star, and the initial pericenter at capture. The pericenter distance of the surviving star has migrated to 10--100 AU, making it potentially the most extreme member of the S-star cluster. From the ejection rate of S5-HVS1-like stars, we estimate that there may currently be a few stars in such highly eccentric orbits. They should be detectable (typically $K_{\rm s}\lesssim 18.5\,$mag) by the GRAVITY instrument and by future  Extremely Large Telescopes and hence provide an extraordinary probe of the spin of SgrA*.

% The probability of the captured star avoiding (secular) tidal disruption and being detectable by the GRAVITY instrument (at $K\lesssim 19\rm\,mag$ \note{[check]}) is $\sim 1/3$. Integrating over the history, the current number of detectable stars in similar orbits near the BH is estimated to be between 5 and $500$. The highly relativistic orbits of these stars, if identified, can be used to measure the BH's spin if $\chi\gtrsim 0.1$. The binary break-up channel contributes a main-sequence EMRI rate $\gtrsim 10^{-7}\rm\, yr^{-1}$.
\end{abstract}


\section{Introduction}
\citet{2020MNRAS.491.2465K} reported a hyper-velocity star (HVS) S5-HVS1 that is confidently associated with the Galactic Center (GC). The HVS is an A-type main-sequence star of about $2.3\msun$ and its inferred ejection speed from the GC is $v_{\rm e}\simeq 1800 \rm\,km\,s^{-1}$.
% The age of the star is loosely constrained to be between 30 and 100$\,$Myr (1-$\sigma$ error).
The ejection speed being much larger than the surface escape speed of the star rules out the ejection scenarios of supernova explosion in a close binary or dynamical encounters between binaries \citep[see e.g.,][]{Per+12}. The most likely mechanism is the tidal break-up of a tight binary system by the supermassive black hole (BH) SgrA* \citep{1988Natur.331..687H}, where one star is ejected to infinity at high speed and the other is bound. The ejection speed is roughly $\ve\sim v_{\rm orb} (M/m_t)^{1/6}$ \citep{2003ApJ...599.1129Y}, where $v_{\rm orb}$ is the internal orbital speed of the binary and $M/m_{\rm t}$ is the BH-to-binary mass ratio, so a supermassive BH is required for the observed ejection speed. Assuming the GC origin, the current location and speed of S5-HVS1 give a travel time of about 5 Myr since ejection \citep{2020MNRAS.491.2465K}.

It has been proposed \citep[e.g.,][]{2003ApJ...592..935G, 2007ApJ...656..709P} that the above Hills mechanism is responsible for the population of both HVSs \citep{brown05_HVS} and the young stars\footnote{See \citet{2005PhR...419...65A} and \citet{2010RvMP...82.3121G} for a discussion of the many proposed formation scenarios for these stars, and \cite{Per09}; \citet{Per+10} for constraints on such origins.}  in the inner 0.04 pc (the \say{S-star cluster}) in roughly isotropic distribution with high eccentricities  \citep{2010RvMP...82.3121G, 2017ApJ...837...30G}. If we make a simple assumption that S5-HVS1 was initially in an equal-mass binary, then the captured star has a specific binding energy of $-\ve^2/2$, which corresponds to an orbit with semi-major axis of about 1000 AU --- similar to the S2 star. Since the eccentricity of the captured star is roughly $1-e\lesssim (m_{\rm t}/M)^{1/3}\sim 0.01$, the pericenter distance is of order 10 AU or less --- a factor of 10 smaller than that of the S2 star \citep{2020arXiv200407187G}. We see that the captured star should be (at least temporarily) an extreme member of the S-star cluster. Its relativistic orbit provides an extraordinary probe of the spin of SgrA* \citep{2018MNRAS.476.3600W}, provided that it survives until today.

However, the captured star may have been tidally disrupted in the past 5 Myr as a result of the following two processes. First, for each pericenter passage, internal stellar oscillation modes are excited by the tidal potential \citep{1977ApJ...213..183P}, and the mode energy may be dissipated into heat as it grows past an instability threshold \citep{1996ApJ...466..946K, 2001ApJ...546..469W, 2012ApJ...751..136W}. This cumulative tidal heating may lead to runaway radius expansion and eventual tidal disruption of the star \citep{2003ApJ...590L..25A, Ant+11, 2013MNRAS.429.3040L}. Additionally, in the N-body environment near SgrA*, the star experiences numerous scattering by other objects (most importantly stellar-mass BHs) in the process of relaxation and the resulting angular momentum diffusion may push the pericenter inwards and cause tidal disruption \citep{2007ApJ...656..709P,Per+09}. 

In this paper, we first provide general constraints on the mass and orbital parameters of the captured star by studying the binary break-up process in \S\ref{sec:break-up}. Then in \S\ref{sec:MC}, we take a statistical approach to obtain the probability distribution of the mass and orbital parameters of the captured star. Taking the results from \S\ref{sec:MC} as initial conditions, we model the interior and orbital evolution of the star for 5 Myr to study its survival likelihood, taking into account tidal heating (\S\ref{sec:tidal_heat}) and orbital relaxation (\S\ref{sec:Nbody}). We discuss the implications and the prospects of future observation in \S\ref{sec:implication}. A summary is provided in \S\ref{sec:summary}. We use the convention $Q = 10^xQ_x$ in CGS units, except that all masses are denoted in units of solar mass $\msun$. We take the mass of SgrA* and its distance to be $M= 4.26\times 10^6\msun$ and $D=8.25\,$kpc \citep{2020arXiv200407187G}.

\section{Binary break-up}\label{sec:break-up}
We consider a binary system of component masses $m_{\rm e}$ (star 1, the ejected one) and $m_{\rm b}$ (star 2, the bound one), with a semimajor axis (SMA)  $a$, initially in a parabolic orbit near the BH with a pericenter distance $\rp$.
% The black hole mass is $M = 10^{6.6}\msun$ \citep{2016ApJ...830...17B}.
More realistically, the initial orbit is hyperbolic with an asymptotic (thermal) speed of a few hundred $\rm km\,s^{-1}$. The initial kinetic energy is much smaller than the energy of the ejected star $\me \ve^2/2$ and is hence negligible for our purpose. For the same reason, we also ignore the binding energy of the initial binary. The original binary orbit is assumed to be circular, because, as we will show later, their initial separation is constrained to be about a few times the size of ejected star and hence their orbits should have been tidally circularized.

\begin{figure}
  \centering
\includegraphics[width = 0.45 \textwidth]{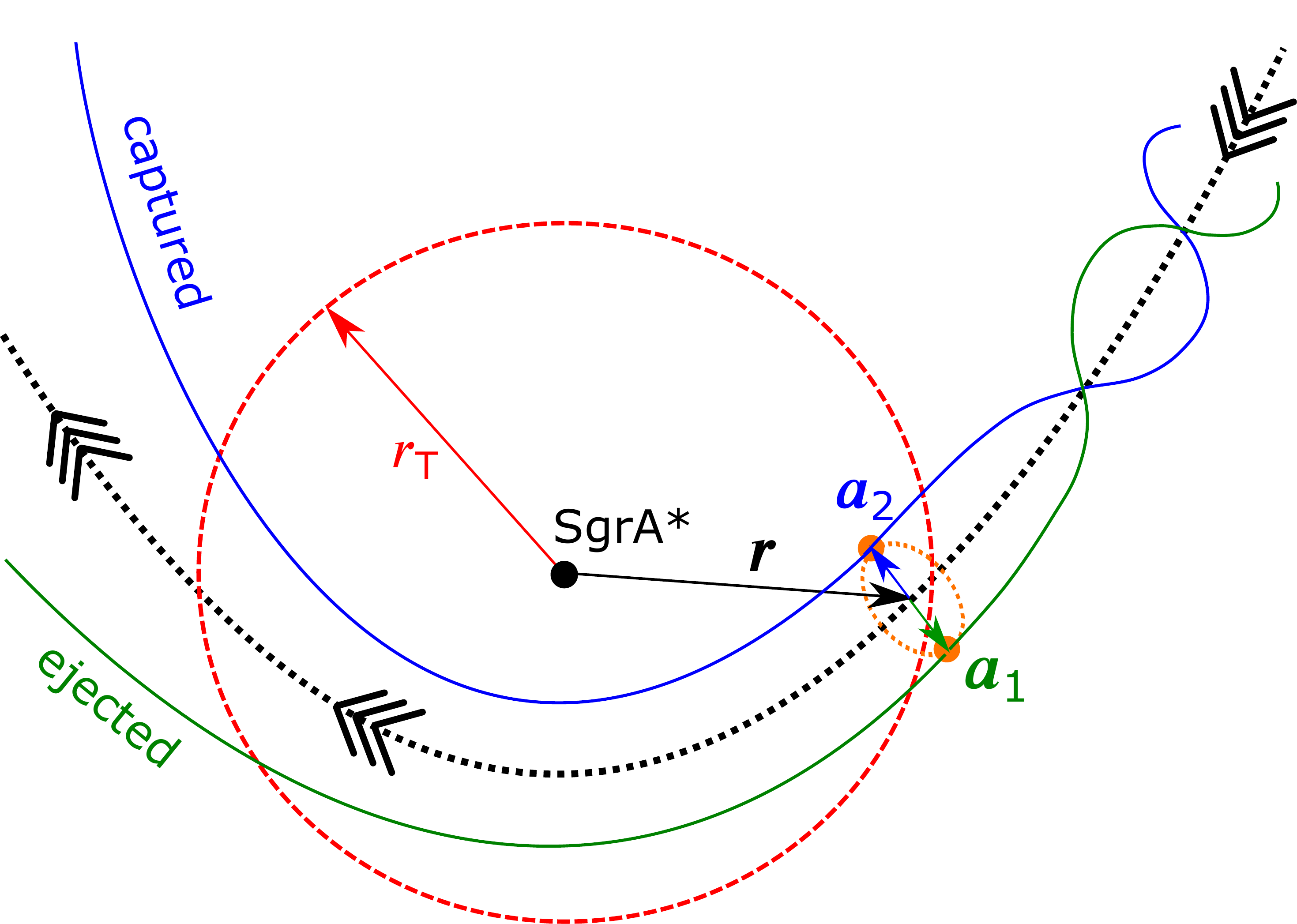}
\caption{Sketch of the binary break-up process (not to scale). The binary center of mass is initially in a nearly parabolic orbit (black dotted line with arrows). When the system reaches below the binary tidal break-up radius $r_{\rm T}$ (red dashed circle), tidal forces from SgrA* tear the two stars apart, and the one closer to the BH (blue trajectory) is captured and the farther one is ejected at high speed (green trajectory). The vector from the BH to the binary center of mass is denoted as $\b{r}$, and the vectors from the binary center of mass to each star are $\b{a}_1$ and $\b{a}_2$. 
}\label{fig:breakup}
\end{figure}

Our coordinate system is centered on the BH, which stays at rest. The binary center of mass position is denoted as $\b{r}$, and each star's position with respect to the center of mass is $\b{a}_1$ and $\b{a}_2$, as shown in Fig. \ref{fig:breakup}. The separation between the two stars is $a = a_1 + a_2$. We define a unit vector $\b{e}_{21}$ pointing from star 2  to star 1, and then the stellar position vectors can be written as
\begin{equation}
  \label{eq:2}
  \b{a}_1 = {\mb  a \over \me  + \mb} \b{e}_{21},\
  \b{a}_2 = -{\me  a \over \me  + \mb}
  \b{e}_{21}.
\end{equation}
The BH's tidal acceleration on star 1 is
\begin{equation}
  \label{eq:3}
  \b{g}_{\rm T,1} = \b{g}(\b{r} + \b{a}_1) - \b{g}(\b{r}) = {GM a_1
    \over r^3} (3\cos\theta \,\hat{\b{r}} - \b{e}_{21}),
\end{equation}
where we have ignored higher order terms and defined
\begin{equation}
  \label{eq:4}
  \cos\theta \equiv \hat{\b{r}}\cdot \b{e}_{21}.
\end{equation}
This should be compared with the gravitational attraction from star 2 on star 1
\begin{equation}
  \label{eq:7}
  \b{g}_{1} = - {G\mb  \over a^2} \b{e}_{21} =- {G(\me  + \mb ) a_1\over
  a^3} \b{e}_{21}.
\end{equation}
We consider the binary to be instantaneously disrupted when $g_1=g_{\rm T,1}$ (equivalent to $g_2=g_{\rm T, 2}$), and this defines the binary tidal break-up radius
\begin{equation}
  \label{eq:8}
  \rT = fa \left(M\over \me  + \mb \right)^{1/3},
\end{equation}
where $f$ is a numerical factor of order unity given by
\begin{equation}
  \label{eq:9}
  f = |3\cos\theta \,\hat{\b{r}} - \b{e}_{21}| =
  (1+3\cos^2\theta)^{1/2}\in [1, 2].
\end{equation}
The impulse approximation is reasonable since the tidal acceleration is a very strong function of distance to the BH such that $g_1\gg g_{\rm T,1}$ at distance $r\gtrsim 1.3\rT$ and $g_1\ll g_{\rm T,1}$ at $r\lesssim 1.3\rT$.

At the binary break-up radius, both stars have nearly the same speed $v_{\rm T} = \sqrt{2GM/\rT}$ or specific kinetic energy $GM/\rT$. Their potential energies are different due to different positions, with star 2's potential energy being $-GM/\rT (1 + a_2\cos\theta/\rT)$ and star 1's being $-GM/\rT (1 - a_1\cos\theta/\rT)$. Our assumption of impulsive disruption means each one's total orbital energy is conserved right after the binary break-up, so we can write the two orbital energies as
\begin{equation}
  \label{eq:10}
  \eps_1 = \me  v_{\rm e}^2/2, 
\end{equation}
\begin{equation}
  \label{eq:11}
  {\eps_2\over \mb } = {GM\over \rT} - {GM\over \rT} (1 +
  {a_2\cos\theta \over 
    \rT}) =  -{GM\over \rT^2} {\me  a\over \me  + \mb } \cos\theta,
\end{equation}
and total energy conservation gives
\begin{equation}
\label{eq:energy_conserve}
     \eps_1 + \eps_2=0.
\end{equation}
Combining eqs. (\ref{eq:8}), (\ref{eq:10}), (\ref{eq:11}), and (\ref{eq:energy_conserve}), we
obtain
\begin{equation}
  \label{eq:12}
 a ={2\sqrt{3}\cos\theta\over f^2} a_{\rm max},\ \ \rT =
 {2\cos\theta\over f} r_{\rm T,max}, 
\end{equation}
where, for convenience, we have defined the maximum allowed binary separation $a$ as
when $\cos\theta = 1/\sqrt{3}$ (and $f = \sqrt{2}$),
\begin{equation}
  \label{eq:23}
  \begin{split}
  a_{\rm max} &= \left(M\over \me  + \mb \right)^{1/3}
  {G\mb \over \ve^2}\\
  &= 3.7\times10^{11}\mr{\,cm}\, {\mb  M_{6.6}^{1/3} \over (m_{\rm b} + m_{\rm e})^{1/3}} v_{\rm e,1800}^{-2}.
  \end{split}
\end{equation}
and the maximum allowed binary break-up radius $\rT$ as when $\cos\theta=1$ (and $f = 2$),
\begin{equation}
  \label{eq:16}
  \begin{split}
   r_{\rm T,max} &= \left(M\over \me  + \mb \right)^{2/3} {G\mb \over \ve^2}\\
    &= 1.0\times10^{14}\mr{\,cm} {\mb M_{6.6}^{2/3}\over(\me  + \mb )^{2/3}} v_{\rm
    k,1800}^{-2}.
  \end{split}
\end{equation}
The binary break-up radius should be compared to the tidal disruption radius of star 1,
\begin{equation}
  \label{eq:15}
  \rTe \equiv R_{\rm e}\left(M\over \me \right)^{1/3}
   \simeq (1.3\times10^{13}\mr{\,cm})\, \me^{1/3}
  M_{6.6}^{1/3},
\end{equation}
where we have adopted a simple mass-radius relation\footnote{Our results are only weakly affected by this approximate relation, because it is only used to set the boundaries of the minimum mass of the captured star and the pericenter distance of the pre-break-up orbit. Disruptions due to cumulative tidal heating are treated more accurately based on numerical modeling of the stellar interior structure (see \S \ref{sec:tidal_heat}). } of $R \simeq 1.2 \rsun (m /\msun)^{2/3}$, appropriate for massive ($\gtrsim 1.5\msun$) main-sequence stars at the inferred age ($\sim$50 Myr) and metallicity ([Fe/H]$\sim0.3$) of S5-HVS1 \citep{2020MNRAS.491.2465K}.  The highly centrally concentrated density profile of massive ($\gtrsim 1.5\msun$) stars requires deeper penetration down to $0.5 \rTe$ to cause major disruption where the star loses $\gtrsim 50\%$ of its mass \citep{2013ApJ...767...25G, 2017A&A...600A.124M, 2019arXiv190708205R}. Since the ejected star survived the binary break-up, we require the pericenter distance of the initial binary to be more than $0.5 \rTe$, i.e.
\begin{equation}
    \rp > 0.5 \rTe.
\end{equation}
We note that the bound star may have been disrupted, and the probability will be quantified in the next section. We also require that the two stars are separated by
\begin{equation}
    a > 2\max(R_{\rm e}, R_{\rm b}),
\end{equation}
so that the radius of the more massive star (with larger radius and lower density) is smaller than the effective radius of its Roche lobe \citep{1983ApJ...268..368E}. The precise lower limit for the binary separation depends on the deformation of the stars near Roche-lobe filling. At even shorter binary separation, the more massive star will fill up its Roche lobe and the system is unstable \citep[the detailed consequence is diverse, e.g.,][]{1998NewA....3..443V}.

% because otherwise 
% We take $R_{\rm e} \approx \rsun (\me/\msun)$ and $R_{\rm b}\approx \rsun (\mb/\msun)$.
% Note that star 2 may be a compact object, and in that case its radius $R_{\rm b}$ does not play a role. 

\begin{figure}
  \centering
\includegraphics[width = 0.45 \textwidth]{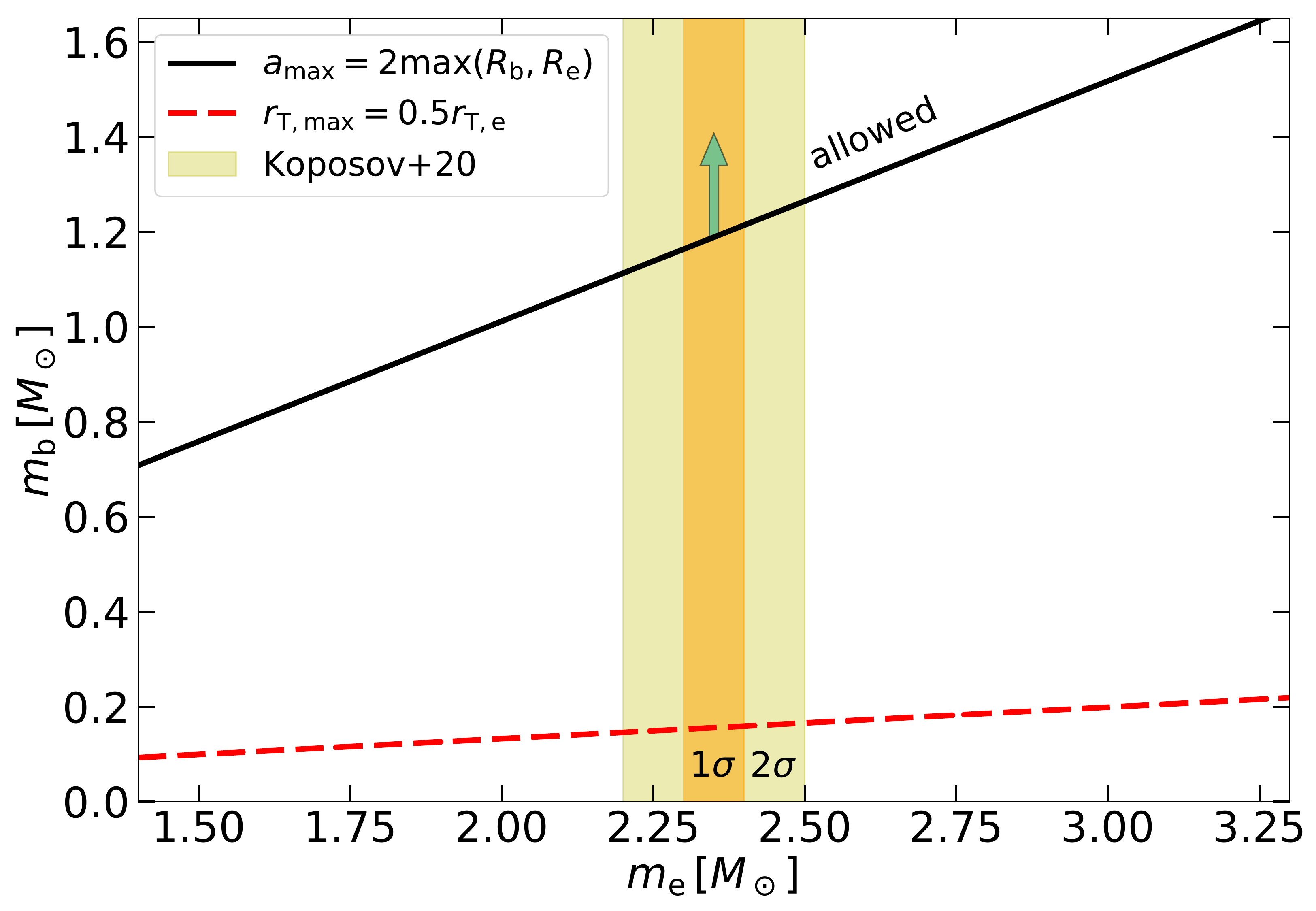}
\caption{The minimum mass of the bound star as constrained by $0.5\rTe < r_{\rm T,max}$ (the ejected star is not tidally disrupted) and $a_{\rm max}>2\max(R_{\rm e}, R_{\rm b})$ (the two stars must be separated before the break-up). The latter is a stronger constraint, so the bound star's mass $\mb$ must be above the solid black line. Shown in yellow is the spectroscopic$+$photometric measurements of $\me $, with $1\sigma$ and $2\sigma$ errors.
}\label{fig:m2min}
\end{figure}

We can rule out very low mass $\mb $ by requiring $0.5\rTe < r_{\rm T,max}$ and $2\max (R_{\rm e}, R_{\rm b}) < a_{\rm max}$, as shown in Fig. \ref{fig:m2min}. We find that $\mb> 1.2\msun$ is generally required. It is possible that the bound star is a massive white dwarf \citep[as the remnant of a massive star $\sim 8\msun$, see][]{2018ApJ...866...21C}, but this requires an additional common-envelope evolution so as to bring the orbital separation down to a few solar radii before the binary break-up. A neutron star or black hole is also possible, but the probability is even smaller, given the steep Salpeter/Kroupa mass function of the progenitor star and that the binary system may be destroyed by the natal kick and mass loss during the supernova. For the above reasons, we only consider the bound star being a main-sequence star.

The SMA of the bound star after binary break-up is denoted as $\ab$ and is given by $\eps_2/\mb  = -GM/2\ab$, i.e.
\begin{equation}
  \label{eq:14}
  \ab = {GM \mb  \over \me  \ve^2} = 1.1\times10^{3}\mr{\,AU}\, {\mb M_{6.6} \over \me} v_{\rm e,1800}^{-2}
\end{equation}
which is similar to that of the S2 star \citep{2017ApJ...837...30G, Do19_S2}.
The Keplerian orbital period is
\begin{equation}\label{eq:Porb}
    P = 18\mr{\,yr} \left(\mb\over \me\right)^{3/2} M_{6.6} v_{\rm e, 1800}^{-3}.
\end{equation}
However, the bound star's pericenter is much closer to the BH than the S2 star (whose $\rp\simeq 120\rm\,AU$). From the binary break-up criterion $\rp\leq \rT$, we know that the orbit of the bound star is highly eccentric $1 - e = \rp/a \sim 10^{-3}$ to $10^{-2}$.

\section{Monte-Carlo Modeling}\label{sec:MC}
In this section, we statistically constrain the pericenter and mass of the captured star. Previous works \citep[e.g.,][]{2006ApJ...653.1194B, 2013ApJ...768..153Z} rely on a large number of 3-body scattering experiments for initial conditions drawn from (assumed) distributions of binary masses, SMAs of the inner binary orbit, and angular momentum of the BH-binary orbit. Our approach is different and simpler in that we make use of all the known information about the ejected star S5-HVS1 instead of drawing random initial conditions blindly.

In the case where the ejected star's kinetic energy is much greater than the initial binding energy of the binary, each member may be ejected at equal probability \citep{2010ApJ...708..605S}. The moment the binary reaches the binary break-up radius, whichever star farther away from the BH is ejected and the closer one is captured, depending on random orbital phase. The inner binary orbit is randomly oriented with respect to the orbit around the BH and the orientation is independent of $\mb $ and the pericenter distance $\rp$.  We denote the probability density function (PDF) of a quantity $x$ as $\mc{P}(x)$ and the corresponding cumulative density function (CDF) is $\mc{F}(x) = \int^x \mc{P}(x') \d x'$. Thus, the joint probability of all three parameters is given by
\begin{equation}
  \label{eq:24}
  \mc{P}(\mb , \rp, \theta) = \mc{P}(\mb )\times \mc{P}(\rp) \times
  \mc{P}(\theta) \times \mc{C}(\mb , \rp, \theta),
\end{equation}
where $\mc{P}(\mb )$ describes the PDF of $\mb $ for a known $\me $ (to be quantified below), $\mc{P}(\rp)$ describes the PDF of the pericenter distance of the initial binary orbit near SgrA* (which in turn depends on the processes scattering the binaries into the SgrA*, e.g. loss-cone dynamics discussed later on), 
%on how the loss cone is filled (to be quantified below), 
$\mc{P}(\theta)$ is the PDF 
of $\theta\equiv \hat{\b{r}}\cdot \b{e}_{21}$ and is considered to be isotropically distributed
\begin{equation}
  \label{eq:25}
  \mc{P}(\theta)\propto \sin\theta,
\end{equation}
and finally $\mc{C}(\mb , \rp, \theta)$ contains the various constraints from known information as summarized below:
\begin{itemize}
\item The bound star's mass is constrained to be $\mb < m_{\rm b,max}=6\msun$ by the age of the binary system before break-up (taken to be $50\rm\,Myr$).
%HBP: Given that the binary is so close, it might have gone through a case A mass-transfer, which could potentially significantly changed the original masses. We should at least briefly address this (annoying, but possible) evolution.

\item The binary break-up radius $\rT$ (and the initial binary separation $a$) is given by eq. (\ref{eq:12}) as required by the measured ejection speed $\ve$.
\item The pericenter radius $\rp\leq \rT$ such that binary break-up occurs.
\item Star 1 is ejected and hence $\theta<\pi/2$.
\item The pericenter radius $\rp>0.5\rTe$ such that the ejected star is not tidally disrupted.
\item The initial binary is separated by $a> 2\max (R_{\rm e},R_{\rm b})$.
\end{itemize}
We note that the order of drawing $\mb $, $\rp$, and $\theta$ is unimportant, since they are independent (except for the constraints above). We reject the entire Monte-Carlo sample of $(\mb , \rp, \theta)$ if any of these constraints is violated and start over. In the following, we describe our choices of $\mc{P}(\mb )$ and $\mc{P}(\rp)$.

We first randomly draw the mass of the bound star $\mb $ according to a power-law binary mass ratio distribution $\mc{P}(q)\propto q^\alpha$ for each particular primary mass, and our fiducial power-law index is $\alpha=0$. This is motivated by statistical studies of short period binaries with mass ratio between 0.3 and 1 by \citet{2017ApJS..230...15M}. We find that our results are only weakly affected for other choices of $\alpha= -0.5$ and $0.5$, because only a small range of mass ratio $0.4\lesssim q<1$ is allowed by the physical constraints of the system. For the same reason, our results are insensitive to possible Case A mass transfer before the binary break-up, which changes the mass ratio by mass transfer between the two stars while both are on the main-sequence.

If the ejected star is the primary ($\mb  < \me $, which occurs half of the time), then we have 
\begin{equation}
  \label{eq:17}
  \mc{P}(\mb ) = {1+\alpha\over 2\me } (\mb /\me )^\alpha  \ \ \
  \mbox{for $\mb  < \me $.}
\end{equation}
If the bound star is the primary ($\mb  > \me $), then we first consider the mass distribution of the primary to be a power-law $\propto \mb ^{-2.3}$ with a Salpeter/Kroupa initial mass function (IMF) slope and then multiply by the probability of mass ratio $q = \me /\mb $, i.e.
\begin{equation}
  \label{eq:18}
  \mc{P}(\mb ) = {0.5(2.3+\alpha)}(\mb/\me)^{-2.3 - (1 + \alpha)} \ \ \  \mbox{for $\mb  > \me $.}
\end{equation}
The CDF combining the two cases is
\begin{equation}
  \label{eq:19}
  \mc{F}(\mb ) = 
  \begin{cases}
    0.5(\mb /\me )^{1 + \alpha}\ \ \  \mbox{for $\mb  < \me $,}\\
    1 - 0.5(\mb /\me )^{-(2.3+\alpha)} \ \ \  \mbox{for $\mb  > \me $.}\\
  \end{cases}
\end{equation}
The random draw of $\mb $ from above is immediately rejected if $r_{\rm T,max}<0.5\rTe$ or $a_{\rm b,max}<2\max(R_{\rm e}, R_{\rm b})$.

The maximum mass of the bound star $m_{\rm b,max}$ has large uncertainty because the age of the binary system is only weakly constrained to be between 30 and 100 Myr \citep[at $1\sigma$,][]{2020MNRAS.491.2465K}. We have tested different choices of $m_{\rm b,max}$ between 5 and $8\msun$ (corresponding to different binary ages) and found the results to be qualitatively similar. A star between $m_{\rm b,max}$ and $\sim8\,\msun$ would have evolved off the main-sequence to become a white dwarf before the binary break-up. Using the initial-final mass relationship from \citet{2018ApJ...866...21C}, we find that the probability of the bound star being a white dwarf with mass above constraint from the minimum binary separation (Fig. \ref{fig:m2min}) is about 1\% in the full loss cone case and even smaller in the empty loss cone case. In reality, this probability should be further reduced, because a common-envelope evolution is require to shrink the binary separation down to a few solar radii. A star $\gtrsim 8\msun$ should have exploded and the binary system may have been destroyed by the mass loss and natal kick during the supernova. We ignore these cases, making a small error given the steepness of the mass function slope. Nevertheless, if the captured star is a compact object, this channel contributes a fraction of extreme mass ratio inspirals (EMRIs).

We also note that the true mass ratio distribution and primary mass function are highly uncertain for tight binaries near the GC, but our approach is general and can be improved when more information is available. For instance, if S5-HVS1 was formed under a top-heavy IMF slope of $-1.7$ as suggested by the population of young massive stars from SgrA* \citep{2013ApJ...764..155L}, then the probability of the captured star being a neutron star or BH may be as large as 5\%. However, our results are only weakly affected if the captured companion of S5-HVS1 is a main-sequence star, the case that we focus on.

% When $\mb $ is successfully drawn, we still need two additional
% parameters: the pericenter distance $\rp$ and
% $\theta=\mr{acos}(\hat{\b{r}}\cdot \b{e}_{21})$. Due to randomness of
% inclination and orbital phase at binary break-up point, it is a
% reasonable approximation to draw $\theta$ from spherically symmetric
% distribution 
% \begin{equation}
%   \label{eq:20}
%   \mc{P}(\theta)= \sin\theta \ \ \ \mbox{for
%     $0 < \theta < \pi/2$,}
% \end{equation}
% where we require $\theta<\pi/2$ since star 1 is
% ejected. We also impose the following restrictions (whichever is more
% stringent): $a>R_1 + R_2\approx \rsun (\me  + \mb )/\msun$ (the  
% initial binary stars do not collide), and $\rT>r_{\rm T*}$ (star 1 is
% not tidally disrupted).

\begin{figure*}
  \centering
\includegraphics[width = 0.85 \textwidth]{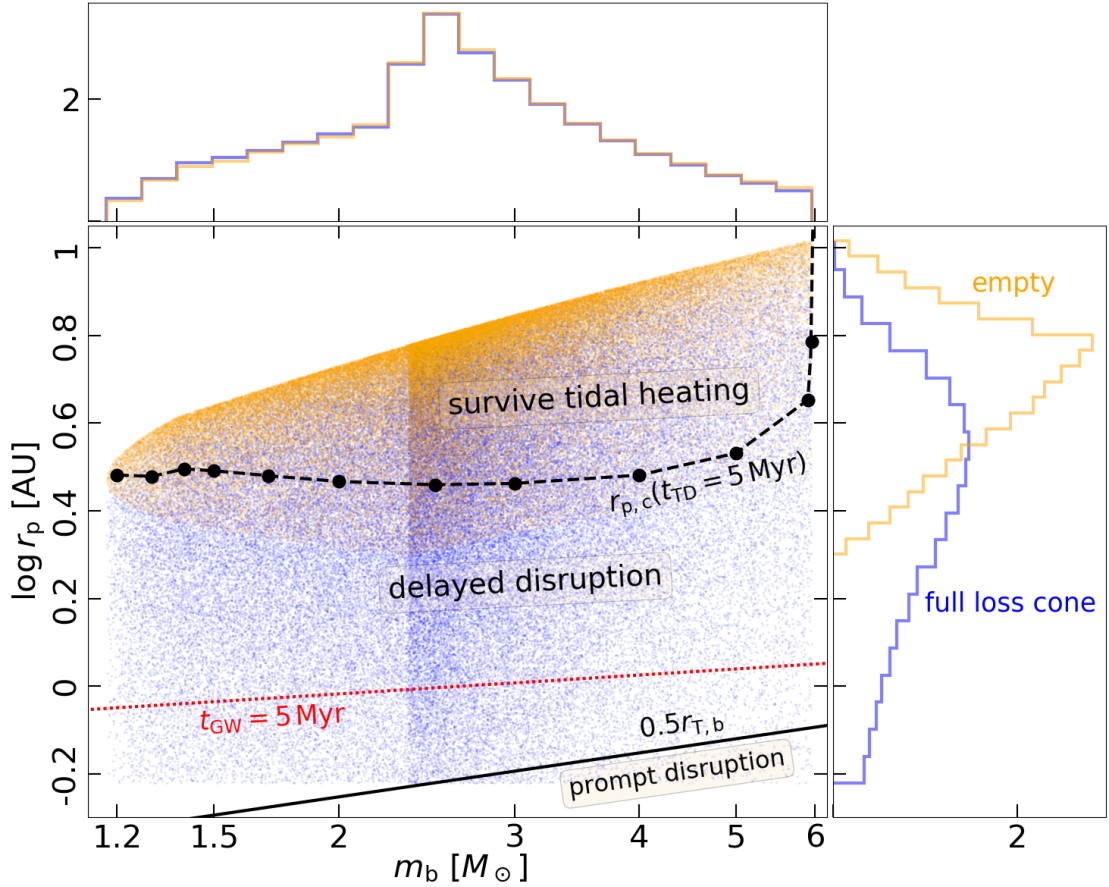}
\caption{Monte-Carlo samples ($N=10^5$) of the mass $\mb $ and pericenter $\rp$ of the bound star. We take the ejected star's mass and ejection velocity to be $\me  = 2.35\msun$ and $\ve=1800\rm\,km\,s^{-1}$, BH mass $M=4.26\times10^6\msun$, and the binary mass-ratio distribution slope $\alpha=0$. The orange and blue samples correspond to empty and full loss cones, respectively. The cutoff towards the top, bottom, left and right sides of the $\mb$-$\rp$ parameter space are due to constraints by the ejection velocity $\ve$, the ejected star's survival of tidal disruption, initial binary being separated at $a> 2\max(R_{\rm e},R_{\rm b})$, and main-sequence age of the ejected star, respectively. The black solid line denotes $0.5r_{\rm T,b}$, below which the bound star is promptly disrupted during the binary break-up. The filled black circles denotes the critical pericenter distance $r_{\rm p,c}$ below which the star is tidally disrupted within 5 Myr after the binary break-up, as a result of cumulative tidal heating (see \S\ref{sec:tidal_heat}). The rapid increase of $r_{\rm p,c}$ above 6$\,\msun$ is because the star evolves off the main-sequence into a red giant. The exact value of main-sequence turn-over mass depends on the age of the pre-break-up binary system, which is taken to be 50 Myr for this example. Stars above the black dashed line survive the tidal heating for at least 5 Myr, provided that their orbits are not strongly modified by scattering off other massive objects (see discussion in \S\ref{sec:Nbody}). The parameters corresponding to gravitational wave (GW) inspiral time $t_{\rm GW}=5\rm \,Myr$ is shown by the red line.
% The detection limit of the GRAVITY instrument for main sequence stars is roughly $\mb \gtrsim 1.7\msun$ ($K\lesssim 19\rm\,mag$), which is shown as a vertical dotted line.
}\label{fig:MC}
\end{figure*}

The pericenter distance distribution $\mc{P}(\rp)$ is more complicated. Typically, binary systems or single stars fall towards the center from large distances near the BH's sphere of influence $\rh=GM/\sigma^2$ where the typical angular momentum is $J_{\rm c}\simeq \sigma \rh$ (for near circular orbits), and $\sigma$ is the velocity dispersion at  $\rh$. The specific angular momentum of the loss cone orbit is 
\begin{equation}
  \label{eq:6}
  J_{\rm lc} = \sqrt{2GM\rT},
\end{equation}
and, for isotropic velocity distribution, only a small fraction $(J_{\rm lc}/J_{\rm c})^2\simeq 2\rT/\rh$ of binary systems have sufficiently high eccentricity as to reach the binary break-up radius $\rT$. These systems are destroyed in one orbital time $P(\rh) \simeq 2\pi \sqrt{\rh^3 /GM}$, during which other systems may be scattered into the loss cone to maintain a quasi-steady break-up rate. A typical system's angular momentum and orbital energy change by order unity in a relaxation time \citep{1987gady.book.....B, 2011PhRvD..84d4024M} 
\begin{equation}
  \label{eq:1}
  t_{\rm rel}(\rh) \simeq 0.2 {M^2 \over \lara{m^2}} {P(\rh) \over N(\rh) \mr{ln}\,\Lambda},
\end{equation}
where $N(\rh)\simeq 2M/\lara{m}$ is the number of stars near $\rh$, $\lara{m}$ and $\lara{m^2}$ are the first and second moments of the stellar mass spectrum, and $\mr{ln}\,\Lambda\sim 10$ is the Coulomb logarithm. The amount of angular momentum diffusion due to 2-body encounters during $P(\rh)$ is $\Delta J^2/J_{\rm c}^2 \simeq P(\rh)/t_{\rm rel}$.

There are two dynamical regimes of loss-cone filling, depending on the ratio between the mean change in angular momentum per orbit $\Delta J$ and $J_{\rm lc}$ \citep{1977ApJ...211..244L}. If $\Delta J \ll J_{\rm lc}$, then the loss cone is largely empty and binary break-up mainly occurs at the boundary for systems with $J\simeq J_{\rm lc}$. The binary break-up rate is given by the rate at which systems diffuse into the loss cone $\sim N(\rh)/t_{\rm rel}$. On the other hand, if $\Delta J \gg  J_{\rm lc}$, the loss cone is fully populated and binary break-up occurs for any angular momentum $J\leq J_{\rm lc}$ (deeply penetrating orbits with $\rp\ll \rT$ are allowed). In this case, the binary break-up rate is $\sim N(\rh) (J_{\rm lc}/J_{\rm c})^2/P(\rh)\simeq N(\rh)\rT/P(\rh) \rh$. For our GC, the loss cone filling depends on the critical ratio
\begin{equation}
  \label{eq:5}
  {\Delta J^2\over J_{\rm lc}^2} \simeq {5 \, \mr{ln}\Lambda \over 2}  \, {\rh\over \rT} {\lara{m^2}\over M\lara{m}}.
\end{equation}
Taking $\rh\simeq 2\rm\,pc$, $\rT\simeq 5\rm\,AU$ (for tight binaries \textit{only}), the mean stellar mass $\lara{m}\simeq 0.5\msun$, the mean squared stellar mass $\lara{m^2}\simeq 1\msun^2$, we find that the loss cone should be full with $\Delta J^2/J_{\rm lc}^2\sim 10$ near $\rh$ for tight binaries considered in this work. Other processes such as resonant relaxation and non-spherical/time-dependent gravitational potential, may modify the angular momentum diffusion rate compared to that given by 2-body relaxation \citep[see][for recent reviews]{2013CQGra..30x4005M, 2017ARA&A..55...17A, stone20}. We note that molecular clouds, and spiral arms near the GC (a few to ten pc) can enhance the rate of relaxation and hence loss-cone filling compared to stellar perturbers. The full loss-cone case is consistent with the number of young HVSs and S-stars in the GC \citep[if the S-stars are mainly from the Hills mechanism,][]{2007ApJ...656..709P,Ham+17}.

In the following, we draw the distribution of pericenter distance $\rp$ by considering two extreme cases of empty and full loss cones. The reality should be somewhere in between, especially considering that binary may come from the nuclear star cluster within the central 0.5 pc from SgrA* \citep[e.g., in the eccentric disk instability scenario,][]{2009ApJ...697L..44M, Generozov20}, and in this case the loss cone has a larger size and may be closer to the empty side. We also note that even when $\Delta J \ll  J_{\rm lc}$, the phase space deeply within the loss-cone can still be populated by rare strong scatterings \citep{2017MNRAS.468.1760W}.

In the empty loss cone case, we first draw $\theta$ from isotropic distribution.
% If any of the constraints is violated, then we start over by drawing $\mb $ (and checking constraints) and then $\theta$ again. 
Then, the pericenter distance $\rp$ is taken to be the largest allowed,
\begin{equation}
  \label{eq:21}
  \rp=\rT(\theta) \ \ \ \mbox{for empty loss cone.}
\end{equation}
%HBP - we should check that the results for the empty loss-cone do not significantly change when considering that the empty loss cone is somewhat less empty than thought - see Sari's recent papers on the subject.
In the full loss cone case, we first draw $\rp$ distribution from
\begin{equation}
  \label{eq:22}
  \mc{P}(\rp) = (r_{\rm T,max} - 0.5\rTe)^{-1} \ \ \ \mbox{for full loss cone,}
\end{equation}
since the probability of reaching down to a distance $<\rp$ is $\int_0^{\rp}\mc{P}(\rp)\d \rp \propto J^2(\rp)\propto \rp$ and hence $\mc{P}(\rp)$ is constant. The allowed range is $0.5\rTe<\rp <  r_{\rm T,max}$ (see eqs. \ref{eq:16} and \ref{eq:15}), so that the ejected star is not tidally disrupted and its ejection speed is $\ve$. Then, we randomly draw $\theta$ from eq. (\ref{eq:25}).
% , and if any of the constraints above, especially $\rp>\rT(\theta)$, is violated, we start over by drawing $\mb $, $\rp$, and then $\theta$ again.

In Fig. \ref{fig:MC}, we show Monte Carlo sampling of the probability distribution of $\rp$ and $\mb $, for two cases of empty and full loss cones. For the empty (full) loss cone case, the median values of these two parameters are $\rp \sim 5\rm\,AU$ ($3\rm\,AU$) and $\mb \sim 2.5\msun$.

The pericenter distance $\rp$ is related to the semimajor axis $\ab$ by $\rp = \ab(1-e)$, and in extreme eccentricity limit $1-e\ll1$, the gravitational wave (GW) inspiral time is
\citep{1964PhRv..136.1224P} 
\begin{equation}
  \label{eq:26}
  t_{\rm GW} = (2.6\mr{\,Myr})\, \mb^{-1}M_{6.6}^{-2} {r_{\rm p,13}^{7/2} a_{\rm b,16}^{1/2}}.
\end{equation}
Since the semimajor axis $\ab$ is determined for given $\mb$ (by eq. \ref{eq:14}), the GW inspiral time of the bound star only depends on the pericenter $\rp$. There is a critical $r_{\rm p,GW}$ that corresponds to an inspiral time of $5\,$Myr (since the binary break-up), given by
\begin{equation}
  \label{eq:13}
  r_{\rm p,GW} = (1.1\times10^{13}\mr{\,cm})\, \left(t_{\rm GW}\over 5\mr{\,Myr}\right)^{2/7}(\me \mb)^{1/7} M_{6.6}^{3/7} v_{\rm e,1800}^{2/7},
\end{equation}
The orbit is strongly affected by GW only if $\rp \lesssim r_{\rm p,GW}$. We find that GW inspiral is typically unimportant for the orbital evolution of the bound star in the past 5 Myr since break-up.

\section{Fate of the captured star}
In this section, we study the fate of the bound star by grouping the Monte Carlo samples into different classes in the $\rp$-$\mb$ plane as follows:
\begin{itemize}
    \item If $\rp < 0.5\rTb$, the star is promptly disrupted.
    \item If $0.5\rTb < \rp < r_{\rm p,c}$, the star experiences strong tidal heating over many orbits, expands, and is then tidally disrupted \citep{Ant+11, 2013MNRAS.429.3040L}. The critical radius $r_{\rm p,c}$ below which cumulative tidal heating leads to expansion and disruption within 5 Myr is calculated in \S\ref{sec:tidal_heat}.
    \item If $\rp> r_{\rm p,c}$, the star is not tidally disrupted within 5 Myr, provided that its orbit is not strongly modified by scattering of other massive objects in the field. In \S\ref{sec:Nbody}, we study the angular momentum diffusion of the star in the N-body environment near SgrA* and calculate the probability that the star is scattered down below $r_{\rm p,c}$.
\end{itemize}

\subsection{Cumulative tidal heating}\label{sec:tidal_heat}
%HBP: I believe you should address the paper by Antonini, Lombardy & Merritt (2011) in this section - I added the ref. to the bib)
For each pericenter passage, energy is injected into the star's normal oscillation modes via tidal excitation according to the linear perturbation theory \citep{1977ApJ...213..183P}
\begin{equation}
    \Delta E = {2\pi^2GM^2\over \Rb} \sum_{l=2,3, ...}\left(\Rb \over r_{\rm p}\right)^{2l+2} \sum_{n,m}|Q_{nl}|^2 |K_{nlm}|^2,
\end{equation}
where $(n, l, m)$ are the quantum numbers for each eigenmode (\say{radial} order $n$ and spherical harmonic indices $l, m$), the coefficient $Q_{nl}$ is a radial overlap integral for the coupling between the (assumed Newtonian) tidal potential to a given mode, and $K_{nlm}$ is a temporal overlap integral representing the coupling to the orbit. For $r_{\rm p}/r_{\rm T,b}\gtrsim 2$, the quadrupole ($l=2$) modes dominate, so we approximate
\begin{equation}
    \Delta \t{E} = {\Delta E\over G\mb^2/\Rb} \approx \chi^{-6} T(\chi, \mb),\ \  \chi\equiv {r_{\rm p}\over r_{\rm T,b}},
\end{equation}
where $\chi$ is the dimensionless pericenter distance (in units of the tidal radius $\rTb$), and for highly eccentric orbits $1-e\ll 1$, the tidal coupling constant $T\equiv 2\pi^2\sum_{n,m}|Q_{nl=2}|^2 |K_{nl=2m}|^2$ can be expressed as a dimensionless function of the stellar mass $\mb$ (controlling the interior structure) and $\chi$, because nearly all the tidal interactions occur near the pericenter.
% The dimensionless tidal energy deposit can be written as
% \begin{equation}
%     \Delta \t{E} = {\Delta E\over G\mb^2/\Rb} \approx 1.4\times10^{-5} \left(\chi\over 3\right)^{-6} {T(\chi, \mb)\over 10^{-3}}.
% \end{equation}
For fixed stellar interior structure, $T(\chi)$ is a rapidly decreasing function for $\chi\gtrsim 1$ \citep{1977ApJ...213..183P, 1986ApJ...310..176L, 1987ApJ...318..261M}.

We follow \citet{2013MNRAS.429.3040L} by assuming that the mode energy is rapidly dissipated as a result of non-linear coupling to a large number of daughter modes in the radiative zone of the star \citep{2012ApJ...751..136W}. This is because gravitational scattering in the dense environment near SgrA* causes stochastic change in orbital period of the captured star, with a fractional change per orbit $\Delta P/P\sim \sqrt{P/t_{\rm rel}}$, where $t_{\rm rel}$ is the local relaxation time. For typical mode frequency $\omega\sim \sqrt{3GM_*/R_*^3}$, we find $\omega \Delta P\sim 30 (P/20\mr{\,yr})^{3/2} (t_{\rm rel}/1\mr{\,Gyr})^{-1/2} (M_*/3\msun)^{-1}\gg 1$ for all cases considered in this work, so the system is in the chaotic tide regime where mode energy can stochastically grow (by $\Delta E_{nlm}$ on average in each orbit) and is then rapidly dissipated by non-linear effects \citep{Mardling95_1, Mardling95_2, 2004MNRAS.347..437I, 2019MNRAS.484.5645V}.

If the star reacts adiabatically and maintains hydrostatic equilibrium, then after each pericenter passage, the stellar radius increases by an amount $\Delta \Rb \sim \Delta \t{E} \Rb$. Cumulative heating over a large number $(\sim 0.1\Delta \t{E}^{-1})$ of orbits will eventually lead to the runaway expansion of the star and hence disruption. In reality, different layers of the star may radiate away the tidally generated heat, especially if the heat is deposited in the outer layers of the star where it can quickly diffuse to the surface.
% \note{[draw a line of $\Delta E/P=L_*$?]}

We study the response of the captured star in realistic orbits using $\mathtt{MESA}$ \citep[version 15140,][]{Paxton2019} simulations. First, we calculate $T(\chi, \mb)$ with $\mathtt{GYRE}$ \citep[version 6.0.1,][]{2013MNRAS.435.3406T} for a grid of $1.5<\chi<5$ and $1.2<\mb<6$, where the interior structure of each star is given by $50\rm\, Myr$ of $\mathtt{MESA}$ evolution from $Z=0.03$ metallicity zero-age main sequence (without tidal heating). Detailed procedures for calculating $T(\chi, \mb)$ are explained in Appendix \ref{sec:tidal_coupling}. Metallicity only weakly modifies the tidal coupling constant for the stellar mass range of interest. Then, during the $5\rm\, Myr$ of further evolution, we add an orbit-averaged tidal heating rate of $\Delta E/P$ uniformly distributed per unit mass in the radiative zone. We take into account the time dependence of $\chi$ (and hence $\Delta E$) due to stellar radius evolution as given by $\mathtt{MESA}$. This approximation ignores the detailed structural responses of the star due to tidal heating, which only leads to a small correction as shown by \citet{2013MNRAS.429.3040L}. In this subsection, we ignore orbital angular momentum diffusion and fix the orbital pericenter $\rp$ and period $P$ throughout the evolution.

\begin{figure}
  \centering
\includegraphics[width = 0.45 \textwidth]{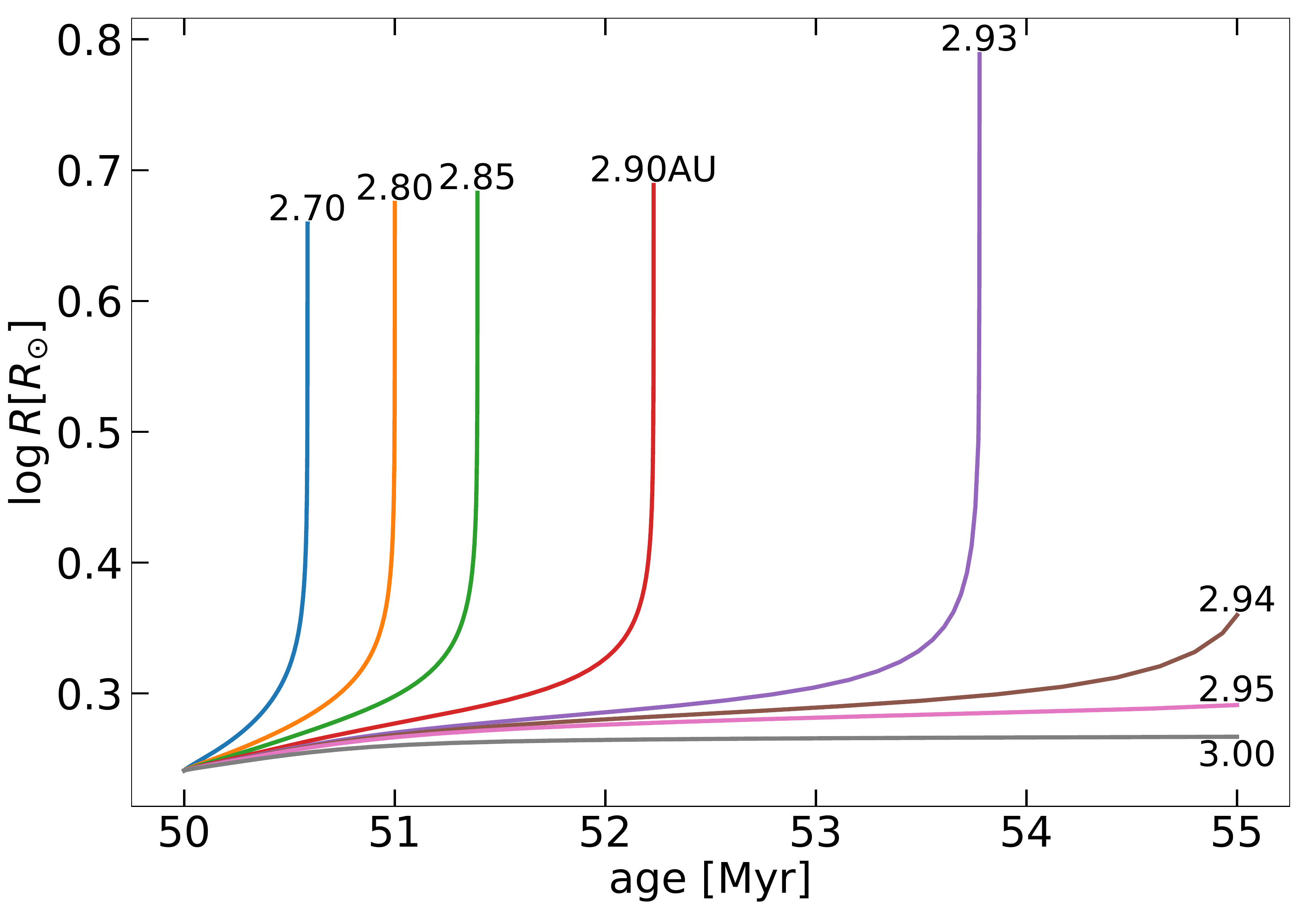}
\caption{Radius evolution from 50 Myr (binary break-up) to 55 Myr (now) for the $\mb=2\,\msun$ case, for different pericenter distances (in units of AU) marked with each line. We take the ejected star's mass and ejection velocity to be $\me  = 2.35\msun$ and $\ve=1800\rm\,km\,s^{-1}$, and the BH mass $M=4.26\times10^6\msun$. The orbital period is $P = 14.5\,$yr as given by eq. (\ref{eq:Porb}). We find that, for $\mb=2\,\msun$, the critical pericenter distance to be $r_{\rm p,c}\approx 2.93\,$AU, below which the star is disrupted within 5 Myr after the tidal capture.
}\label{fig:tidal_heat}
\end{figure}

We show the radius evolution of the star as a function of time, for $\mb=2$ and different pericenter distances, in Fig. \ref{fig:tidal_heat}. Similar experiments are done for other stellar masses and pericenter distances, and in this way, we find the critical pericenter distance $r_{\rm p,c}$ below which cumulative tidal heating leads to expansion and disruption within 5 Myr after the tidal capture. The critical pericenter $r_{\rm p,c}$ as a function of $\mb$ is shown in Fig. \ref{fig:MC}. We find $r_{\rm p,c}\simeq 3\,$AU (roughly independent of the stellar mass) which corresponds to about $2\rTb$ instead of $4\rTb$ as suggested by \citet{2013MNRAS.429.3040L}. The difference is because (1) the orbital periods in our cases are longer (as a result of higher eccentricity) allowing more cooling time, and (2) the stars between 1.5 and $6\,\msun$ have more centrally concentrated density profiles \citep[than the 1 and $20\,\msun$ cases considered by][]{2013MNRAS.429.3040L} and are less susceptible to tidal heating. In the next subsection, we study the angular momentum diffusion of the captured star in the N-body environment near SgrA* under the (conservative) assumption that the star is disrupted if its pericenter is driven to be below $r_{\rm p,c}$.

It should be noted that the age of the binary system is only loosely constrained to be between 30 and 100 Myr (1$\sigma$ confidence interval). This uncertainty affects the radius and interior structure of the captured star especially if it is near the main-sequence turnoff. The main effect is that there is a maximum mass for the captured star $m_{\rm b,max}\simeq 8$, $6$ (our fiducial case), $4.5\msun$ for the age of 30, 50 (fiducial), 100 Myr, respectively. Doing similar experiments for an age of 30 or 100 Myr, we generally find $r_{\rm p,c}\simeq 3\,$AU for $\mb < m_{\rm b,max}$. As can be seen from Fig. \ref{fig:MC}, the consequence of different $m_{\rm b,max}$ is that the captured star has lower mass and closer pericenter distance if we choose an older age. Since the survival probability is only weakly dependent on the mass of the bound star, the uncertainly due to binary age is subdominant compared to that from the relaxation time near SgrA*, which strongly affects the angular momentum diffusion of the post-capture orbit (see the next subsection).

\subsection{Angular momentum diffusion}\label{sec:Nbody}
In this subsection, we consider the orbital evolution of the captured star in the N-body environment near SgrA*.

Unlike the quasi circular orbit case where the scattering timescales in logarithmic orbital energy and angular momentum are comparable and equal to the relaxation time $t_{\rm rel}$, here the newly captured star is in highly eccentric orbit with angular momentum $J\ll J_{\rm c}=\sqrt{GM/\ab}$, so the timescale for changing $\mr{log}\,J$ by unity, $t_{\rm J}\sim (J/J_{\rm c})^2 t_{\rm rel}$, is much shorter. The quasi-steady state of a mass segregated cusp near SgrA* can be described by a power-law density distribution $n_*\propto r^{-\alpha_*}$ with a mass-dependent slope that is steeper for heavier components \citep{1977ApJ...216..883B, 2009ApJ...697.1861A, 2017ARA&A..55...17A}. The scattering rate crucially depends on the degree of mass segregation, especially the number of heavier stellar-mass BHs\footnote{Many works found that there may be $10^2$ to $10^3$ stellar-mass BHs within 10 mpc of the GC \citep[e.g.,][]{2006ApJ...649...91F, 2009ApJ...697.1861A, 2010ApJ...708L..42P, 2016ApJ...830L...1A, 2018MNRAS.478.4030G}.} within 10 milliparsec (mpc) from SgrA*, since the relaxation rate is proportional to $\lara{m^2} = \int \d m\, m^2 (\d N/\d m)$ \citep{1987gady.book.....B}.

\begin{figure}
  \centering
\includegraphics[width = 0.43 \textwidth]{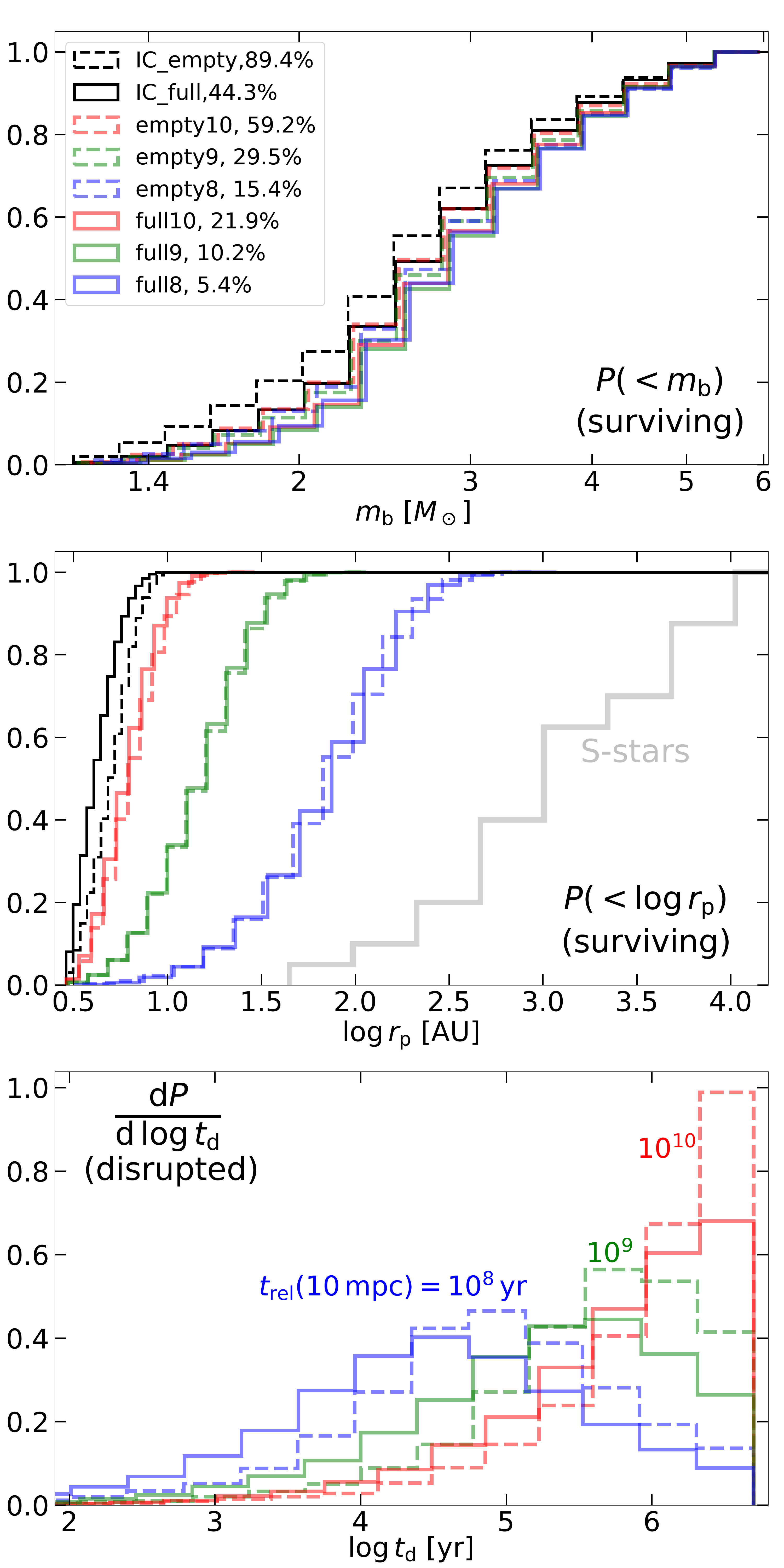}
\caption{The mass (upper) and pericenter (middle panel) distributions for the cases that survive the tidal interactions after 5 Myr of orbital evolution. The initial conditions (ICs), drawn from Fig. \ref{fig:MC}, are shown by black solid and dashed lines for the full and empty loss-cone cases, respectively. The red, green, blue histograms are for the three cases of relaxation timescales $t_{\rm rel}(10\mr{\,mpc})=10^{10}, 10^{9}, 10^{8}\,$yr (see eq. \ref{eq:trel_norm}), respectively. The final surviving percentage in each case is marked in the upper panel's legend. For shorter relaxation time, angular momentum diffusion is faster, more stars are tidally disrupted, and the pericenter distribution of the surviving cases is broader and centered at larger values. The surviving cases typically have pericenters smaller than most of the known S-stars as listed by \citet{2017ApJ...837...30G}, which are shown by the silver line. In the mass distribution of the surviving cases, there is a weak preference for massive ones, mainly because massive stars typically have larger initial pericenters that are farther above $r_{\rm p,c}$. The bottom panel shows the disruption time $t_{\rm d}$ at which the pericenter falls below $r_{\rm p,c}$ (and we assume the star is instantaneously disrupted). All histograms are normalized as probability density (total area below each curve is unity).
% \note{[mark the relaxation timescales in the bottom panel]}
% The mass (upper) and pericenter (lower panel) distributions for the cases that survive the tidal interactions after 5 Myr of orbital evolution. The initial conditions, drawn from Fig. \ref{fig:MC}, are shown by black solid and dashed lines for the full and empty loss-cone cases, respectively. The red, green, blue histograms are for the three cases of field object distributions of BW, NSC1, NSC2 (see text for descriptions), respectively. The surviving percentage in each case is marked in the lower panel's legend. For larger number density of stellar-mass BH increases and more top-heavy mass spectrum, angular momentum diffusion is faster, gives broader pericenter (or eccentricity) distribution, and more stars are tidally disrupted. The pericenter distances of the surviving cases are typically smaller than that of the known S-stars \citep{2009ApJ...692.1075G, 2017ApJ...837...30G}, which is shown by the silver line. In the mass distribution of the surviving cases, there is a weak preference for massive ones, because low-mass stars are more strongly scattered. All histograms are normalized as probability density (total area below the curve is unity).
}\label{fig:hist}
\end{figure}

\renewcommand{\arraystretch}{1.2}
\begin{table*}
\centering
\caption{The probability corresponding to the different fate of the captured star for the empty and full loss cone cases as shown in Fig. \ref{fig:MC}. We consider three choices of relaxation time normalization $t_{\rm rel}(10\mr{\,mpc})=10^{10}, 10^{9}, 10^{8}\,$yr (see eq. \ref{eq:trel_norm}), which depends on the (uncertain) outcome of mass segregation near SgrA*. The star (1) is promptly disrupted if $\rp < 0.5\rTb$ at binary break-up, (2) is disrupted after some delay (of less than 5 Myr) if $0.5 \rTb < \rp < r_{\rm p,c}$, (3) survives until today if $\rp> r_{\rm p,c}$ holds throughout the angular momentum diffusion over the past 5 Myr. Generally, the final survival probability decreases for shorter relaxation time, because a larger fraction of cases are (randomly) scattered to lower angular momentum orbits and are disrupted.
}
\label{tab:probability}
\begin{tabular}{|c|c|c|c|c|} 
\hline
\multirow{2}{*}{loss cone}     & \multirow{2}{*}{$t_{\rm rel}(10\mr{\,mpc})\,\rm [yr]$}         &   prompt disruption (\%) & delayed disruption (\%) & survive (\%) \\ 
\multicolumn{2}{c}{} &  ($\rp<0.5 \rTb$)    &   ($0.5 \rTb < \rp < r_{\rm p,c}$)   & ($\rp > r_{\rm p,c}$) \\
\hline
\multirow{3}{*}{empty} & $10^{10}$ & 0    &   40.8   & 59.2\\

\multicolumn{1}{c}{} & $10^{9}$ & 0    &   70.5   & 29.5\\

\multicolumn{1}{c}{} & $10^{8}$ & 0    &   84.6   & 15.4\\
\hline
\multirow{3}{*}{full} & $10^{10}$ & 0.6    &   77.5   & 21.9\\

\multicolumn{1}{c}{} & $10^{9}$ & 0.6    &   89.2   & 10.2\\

\multicolumn{1}{c}{} & $10^{8}$ & 0.6    &   94.0   & 5.4\\
\hline
\end{tabular}
\end{table*}
\renewcommand{\arraystretch}{1}

We carry out Monte Carlo (MC) scattering\footnote{Resonant relaxation \citep[RR,][]{1996NewA....1..149R} plays an important role for stars with much larger pericenter distances (e.g., the S-stars) but is strongly suppressed by rapid GR apsidal precession for the highly eccentric orbits considered here \citep{2006ApJ...645.1152H, 2011ApJ...738...99M, 2016ApJ...820..129B}. RR is absent in our MC scattering experiment because the angular momentum change in each time step is due to uncorrelated 2-body interactions. } simulations to track 5 Myr of angular momentum diffusion for the cases that survive direct tidal heating with $\rp > r_{\rm p,c}$. This is done using a simplified version of MC method by \citet{1978ApJ...225..603S}, see also \citet{1972ApJ...173..529S, 1977ApJ...211..244L, 1995ApJ...445L...7H, 2005ApJ...629..362H, 2016ApJ...820..129B}. For each orbit, the angular momentum is perturbed by a drift term $\Delta_1 J = J_{\rm c}^2 P/(2t_{\rm rel}J)$ and a stochastic term $\Delta_2J = \pm (P/t_{\rm rel})^{1/2} J_{\rm c}$ (random choice between $+$ and $-$), where $P$ is the orbital period and $t_{\rm rel}(a)$ is the relaxation time at a distance of the SMA $a$ from the BH. This is because in the $J/J_{\rm c}\ll 1$ limit, the orbit undergoes random walk in 2-dimensional J-space with step length $\sqrt{\Delta J^2}$, and geometrically, one can show that the orbit-averaged mean change or drift in angular momentum is related to the mean squared change by $\Delta J = \Delta J^2/2J$ \citep{1977ApJ...211..244L}. Since $(P/t_{\rm rel})^{1/2} J_{\rm c}/J\ll 1$, the stochastic term $\Delta_2J$ is much larger than the drift term $\Delta_1J$, but both are important for timescales of order $(J/J_{\rm c})^2t_{\rm rel}$ or longer.

The GW inspiral time for $r_{\rm p}\gtrsim 3\rm\, AU$ is longer than $100\,$Myr and hence we ignore the angular momentum and energy carried away by GW. We also ignore energy perturbation by scattering because the (energy) relaxation time is much longer than 5 Myr for all cases considered here, so the orbital energy and SMA stay fixed. For $J/J_{\rm c}\ll 1$, dynamical friction only contributes to change in energy on a timescale $(\lara{m}/\mb) t_{\rm rel} \sim t_{\rm rel}$ and is hence ignored. %HBP: why is that? we should either explain or give a reference.

For a density distribution $n_*\propto r^{-\alpha_*}$, we obtain the relaxation time profile $t_{\rm rel}(r)\propto r^{\alpha_*-3/2}/\lara{m_*^2}$, which increases with radius (since mass segregation gives $\alpha_*>3/2$ and $\lara{m_*^2}$ generally increases towards the supermassive BH). We consider the following parametrized power-law
\begin{equation}\label{eq:trel_norm}
    t_{\rm rel} = t_{\rm rel,0} (r/r_0)^{\alpha_{\rm t}},
\end{equation}
normalized at $r_0=10\rm\,$mpc (near the initial SMA of the captured star) and we take $\alpha_{\rm t} = 0.3$ roughly given by the \citet[][BW]{1977ApJ...216..883B} profile for the heaviest objects $\alpha_*=7/4$ (appropriate for the innermost cusp where heavy BHs dominate relaxation). Since the semimajor axis in our MC sample in Fig. \ref{fig:MC} only spans a factor of a few (see eq. \ref{eq:14}), our results only depend weakly on the power-law index $\alpha_{\rm t}$. However, the normalization $t_{\rm rel,0}$ is highly uncertain. In the case of the classical BW weak segregation (when stellar-mass BHs are relatively common), $t_{\rm rel, 0}$ is a few to 10 Gyr. On the other hand, \citet{2009ApJ...697.1861A} show that, if the nuclear star cluster contains only a small fraction of stellar-mass BHs such that relaxation is dominated by lighter objects, then these BHs sink to the center (by dynamical friction) and form a much steeper cusp. In this strongly segregated solution, the relaxation rate within 10's of mpc from SgrA* may be enhanced by one or two orders of magnitude due to larger density of stellar-mass BHs. Above considerations motivate us to consider three different cases of $t_{\rm rel,0} = 10^{10},\ 10^{9},\ 10^{8}\,$yr. The shortest $t_{\rm rel,0}$ roughly corresponds to the strongly segregated model obtained by \citet[][right panel of their Figure 1]{2016ApJ...830L...1A}, which includes two populations of 10 and 30$\,\msun$ BHs (and the latter dominates relaxation).

% ...\note{[Relativistic orbit of the target star is tracked by taking SgrA* to be a Schwarzschild BH. The background star distribution is fixed. ]}

% The first case is the weakly segregated \citet[][BW]{1977ApJ...216..883B} cusp solution with constant $t_{\rm rel}$.

% We consider two difference general cases of weakly segregated   and strongly segregated nuclear star cluster (NSC) solutions by \citet{2009ApJ...697.1861A}. We consider two NSC solutions with different stellar-mass BH mass spectrum \note{[more descriptions here...]} NSC1 and NSC2 stand for the cases where relaxation is dominated by stellar-mass BHs with 10 and 30$\,\msun$, respectively.

For each initial condition (drawn from Fig. \ref{fig:MC}), we evolve the orbit for 5 Myr during which the star is assumed to be rapidly disrupted if the pericenter distance falls below $r_{\rm p,c}$. The simulation results are shown in Fig. \ref{fig:hist}. As expected, a large fraction of the cases are scattered to higher eccentricities and then disrupted. At the same time, some are scattered to lower eccentricities and survive. For both the empty and full loss cone cases, the pericenter distributions of the surviving stars are similar.

The final survival probabilities for all cases are summarized in Table \ref{tab:probability}. We find that the final survival probability of the captured star is between $5\%$ (for full loss cone and fastest relaxation) and $50\%$ (for empty loss cone and slowest relaxation). Of the surviving cases, $\gtrsim 90\%$ are more massive than $2\msun$, which is mainly due to initial conditions rather than the slight preference for higher-mass stars to survive.

In Table \ref{tab:S-stars}, we compared the expected properties of the surviving star to the three known S-stars (S2, S62, and S175) with the closest pericenters $\lesssim 100\,$AU, which are seemingly consistent with our prediction for $t_{\rm rel}(10\mr{\,mpc})\ll 1\,$Gyr. The Hills mechanism predicts that the mass and ejection speed of the ejected star are related to the mass and period of the bound star by eq. (\ref{eq:Porb}), $\me v_{\rm e, 1800}^2 \simeq 1.5\, \mb (P/10\mr{\,yr})^{-2/3}$. According to this relation, we find that none of the three stars was associated with S5-HVS1: S2 is too massive/young, the estimated mass of S62 is too large for its SMA, and the estimated mass of S175 is too large for its SMA. However, S62 could be associated with a different HVS with $\me v_{\rm e, 1800}^2\simeq 7.5$ (slightly more massive and/or faster than S5-HVS1), and S175 could be associated with an HVS with $\me v_{\rm e, 1800}^2\simeq 1.2$ (slightly less massive and/or slower than S5-HVS1). The method developed in this work can generally applied to link future HVSs to S-stars and vice versa.

\section{Implications and future observation}\label{sec:implication}

In this subsection, we discuss the implications on (1) the contribution to tidal disruption event (TDE) rate by binary break-ups, (2) the cumulative number of surviving captured stars within 10 mpc, and (3) using the relativistic orbits of the captured stars to measure the spin of SgrA*.

The Southern Stellar Stream Spectroscopy Survey \citep[$\rm S^5$,][]{2019MNRAS.490.3508L} had covered a small fraction of the sky $\Delta \Omega/4\pi \simeq 330\rm\,deg^2/4\pi \ll 1$.
% The main-sequence lifetime ($\sim1\rm\,Gyr$) and current age ($\sim 50\rm \,Myr$) of S5-HVS1 are both much longer than the time since ejection ($5\rm\,Myr$). Thus,
There are potentially many more S5-HVS1-like stars in other areas of the sky and ejected earlier in time.
% The ejection rate of S5-HVS1-like stars is highly uncertain.
The total ejection rate of HVSs in the mass range of 2.5 to 4$\msun$ is estimated to be about $10^{-6}\rm\, yr^{-1}$ from recent observations \citep{2014ApJ...787...89B, brown18_HVS_rate}. These stars typically have much lower ejection speeds $\sim600\rm\, km\, s^{-1}$. Assuming a flat logarithmic binary separation distribution, \citet{2014ApJ...795..125R} predicted that $\gtrsim 20\%$ (both full and empty loss cones) of HVSs should have speed around $1800\rm\,km\,s^{-1}$. Combining these two arguments gives about one S5-HVS1-like star every 5 Myr. This is in slight tension with the single detection by the $\rm S^5$ survey, which only probed $\sim 1\%$ of the spherical volume.
% \footnote{The $\rm S^5$ survey may be incomplete even within the covered area, because only $\sim4\times10^4$ individual target stars had been observed by the time S5-HVS1 was discovered \citep{2020MNRAS.491.2465K}. }.
This issue may be due to potentially anisotropic angular distribution of ejected stars, non-steady ejection rate, observational biases, and Poisson fluctuation. 

% On the other hand, the \citet{2019MNRAS.490.3508L} survey only probed $\sim 1\%$ of the spherical volume, from which we can infer the total number to be between 1 (the most conservative) and $\sim$100 (most optimistic) per 5 Myr, but with large Poisson error.

In the following, we adopt a fiducial ejection rate of $\Gamma=10^{-6}\Gamma_{-6}\rm\, yr^{-1}$ for S5-HVS1-like stars. Future observations, e.g., \textit{Gaia} DR3 \citep{2018A&A...616A...1G} and Dark Energy Spectroscopic Instrument \citep{2016arXiv161100036D}, will better measure this rate.
% Note that the ejection rate of lower mass stars could be larger by a factor of order unity (but low-mass captured stars are easier to tidally disrupt).
Along with each ejection, there is a captured star. Very few cases are promptly disrupted. However, an order-unity fraction of the captured stars are disrupted after some delay as a result of cumulative tidal heating and angular momentum diffusion. Thus, the binary break-up channel provides a lower limit on the total TDE rate in the Milky Way to be $\gtrsim 5\times 10^{-7}\Gamma_{-6}\rm\, yr^{-1}$ (taking the most conservative case of 50\% disruption). In these eccentric TDEs, the final full disruption is preceded by a few partial disruption events separated by about 20 years \citep[similar to the white dwarf TDE cases considered by e.g.,][]{2014ApJ...794....9M, 2017MNRAS.468.2296V}. The partial TDEs feed the BH at sub-Eddington rate and the accretion disk generates bright X-ray emission. Then, in the full disruption, about half of the star is fed to the BH \citep[see][for a discussion of TDEs from bound stars]{2018ApJ...855..129H}, and both optical and X-ray emission may be generated near the Eddington luminosity \citep[e.g.,][]{2009MNRAS.400.2070S, 2011MNRAS.410..359L, 2016MNRAS.461..948M, 2018ApJ...859L..20D, 2020MNRAS.492..686L}. Another possible but much less likely outcome is that, if the relaxation time is extremely long ($>10\rm\,Gyr$) such that angular momentum diffusion is slower than orbital circularization due to GW emission (such a situation may be realized for larger BH masses $\gtrsim 10^{7.5}\msun$), then the captured star undergoes EMRI \citep[see][for a discussion]{metzger17_EMRI}.

The cumulative number of surviving bound stars in the past 5 Myr is between $0.25\Gamma_{-6}$ (for 5\% survival fraction) and $2.5\Gamma_{-6}$ (for 50\% survival fraction). This is a lower limit because we do not take into account contributions from other lower mass binary systems and captures that occurred more than 5 Myr ago (those systems may have a lower but non-zero surviving probability, depending on the relaxation time).
% Thus, it is reasonable to expect a few surviving, but so far undetected, captured stars.

\begin{figure}
  \centering
\includegraphics[width = 0.45 \textwidth]{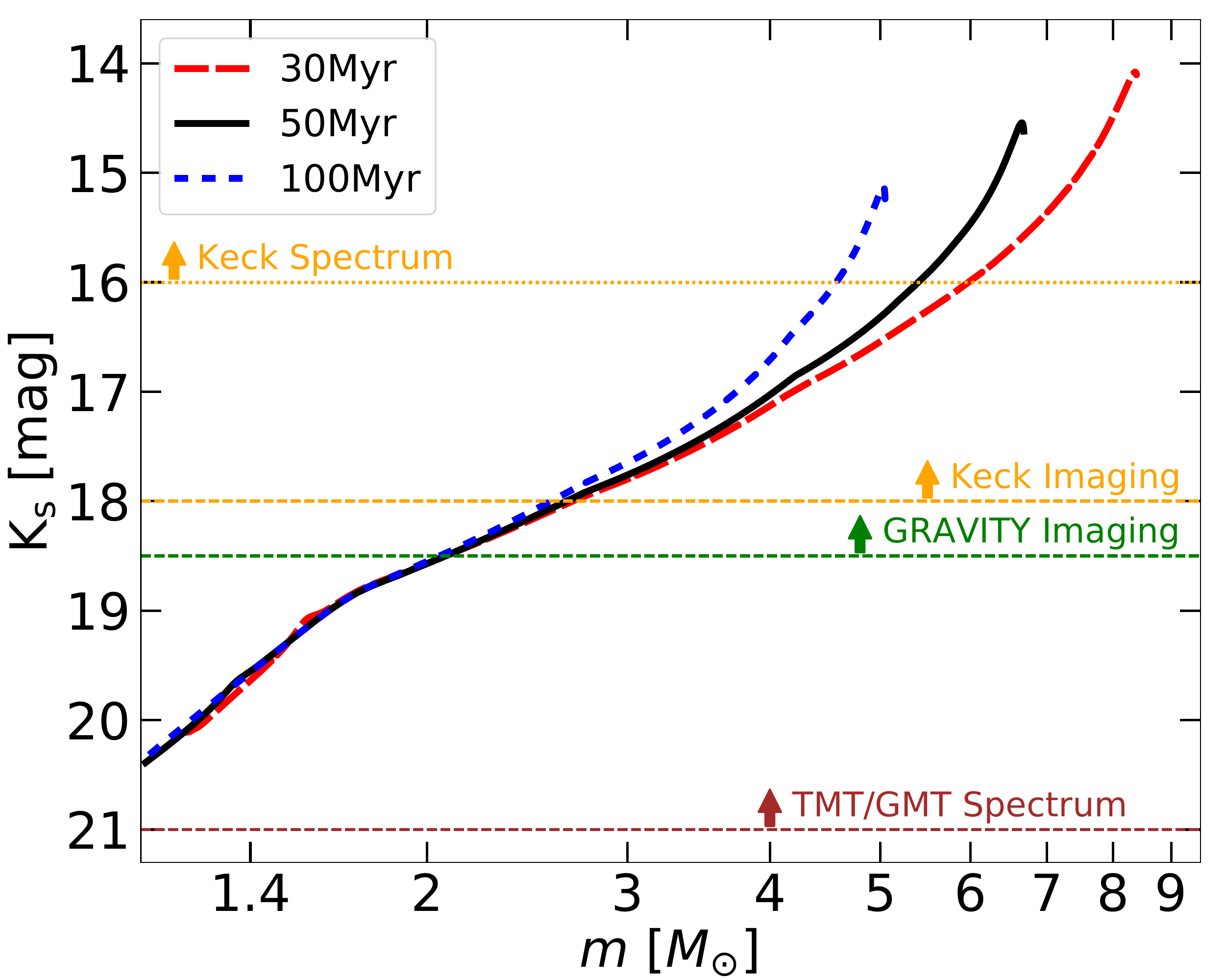}
\caption{$\rm K_{\rm s}$-band magnitude for a stellar population at different ages of 30, 50, 100 Myr. We take metallicity [Fe/H]=0.3 \citep{2020MNRAS.491.2465K} and extinction $A_{\rm Ks}=2.42\,$mag \citep{2011ApJ...737...73F}. Detection limits for various instruments, as estimated by \citet{2017arXiv171106389D}, are shown in horizontal lines including Keck, GRAVITY, and future Giant Magellan Telescope (GMT) and Thirty Meter Telescope (TMT).
}\label{fig:Ks_mag}
\end{figure}

In Fig. \ref{fig:Ks_mag}, we show the estimated Keck NIRC2 (2nd generation near-infrared camera) $\rm K_{\rm s}$-band magnitude of a stellar population at different assumed ages of 30, 50, 100 Myr for metallicity [Fe/H]=0.3. The isochrones are generated by the public code $\mathtt{SPISEA}$\footnote{\href{https://spisea.readthedocs.io/en/latest/}{https://spisea.readthedocs.io}} \citep{hosek20_spisea}, using MESA Isochrones and Stellar Tracks (MIST) stellar evolution \citep{2016ApJ...823..102C} and ATLAS atmosphere models \citep{2003IAUS..210P.A20C}. We adopt extinction $A_{\rm Ks}=2.42\,$mag as estimated by \citet{2011ApJ...737...73F}, see also \citet{2010A&A...511A..18S}. The majority ($\gtrsim 90\%$) of the surviving cases should be brighter than the imaging limit (18.5 mag) for the GRAVITY instrument at Very Large Telescope Interferometer \citep{2020arXiv200407187G}. We conclude that observations in the near future will likely find the captured companions of S5-HVS1-like stars.
% On the other hand, non-detection will provide interesting constraints on the the relaxation time ($t_{\rm rel,0}\lesssim 1\rm\, Gyr$) and hence the number of stellar-mass BHs $N_{\rm bh}\gtrsim 30$ within $10\,$mpc from SgrA*.

If one of the captured stars considered in this work survived till today, it is most likely an undetected extreme member of the S-star cluster with pericenter distance in the range 10 to 100 AU. Its relativistic orbit provides an extraordinary probe of the spin of SgrA*. The Lense-Thirring (LT) precession per orbit is $\Phi_{\rm LT} \simeq 2.2\times10^{-4}\mr{\,rad}\, (\rp/30\mr{\,AU})^{-3/2} \chi \sin i$, 
% $\Phi_{\rm LT}\approx 4\pi \chi (\rg/2\rp)^{3/2}\sin i \simeq 4\times10^{-4}\mr{\,rad} (\rp/500\rg)^{-3/2} \chi \sin i$, 
where $\chi$ is the dimensionless spin of the BH and $\sin i$ is the spin-orbit inclination. Taking a potential astrometric precision of $10\,\rm \mu as \simeq 0.08\rm\,AU$ in projected distance, we see that the GRAVITY instrument is sensitive to BH spin provided that $\chi \sin i \gtrsim 0.3 (\rp /30\,\rm AU)^{3/2}$ \citep[see][for a more detailed calculation]{2018MNRAS.476.3600W}.

\renewcommand{\arraystretch}{1.2}
\begin{table*}
\centering
\caption{Comparison between the properties of the captured companion of S5-HVS1 (\textit{if surviving}) and the three known S-stars with pericenters closest to SgrA*: S62 \citep{Peissker20}, S2 \citep{2020arXiv200407187G}, and S175 \citep[also known as S0-104,][]{Meyer12, 2017ApJ...837...30G}. The SMA and period of the captured star are given by eqs. (\ref{eq:14}) and (\ref{eq:Porb}), respectively, and the pericenter has been modified by angular momentum diffusion over the past 5 Myr (see Fig. \ref{fig:hist}). We find that none of the three stars are consistent with the captured star studied in this work: S2 is too massive/young, S62 has too small SMA for its estimated mass, and S175 has too large SMA for its estimated mass. However, modest variation in the parameters (mass, speed, and age) of the ejected star can be made to fit these three S-stars, which means that they are consistent with the Hills mechanism origin.
}
\label{tab:S-stars}
\begin{tabular}{|c|c|c|c|c|} 
\hline
\multirow{2}{*}{properties}     & {captured companion}         &  \multirow{2}{*}{S62} & \multirow{2}{*}{S2} & \multirow{2}{*}{S175} \\ 
\multicolumn{1}{c}{} &  of S5-HVS1   &   \multicolumn{3}{c}{} \\
\hline
SMA [AU]      & 600--3000     & 740       & 1030      & 3100--3800\\
$P$ [yr]      & 8--90         & 9.9       & 16        & 90--100\\
$\rp$ [AU]    & 5--200        & 18        & 120       & 35--50\\
K-band magnitude & 19.5--15.5    & $\sim$16  & 14        & 17.5\\
mass [$\msun$]   & 1.4--6        & $\sim$5   & $\gtrsim$10 & $\sim$3.5\\
\hline
consistent with     & \multirow{2}{*}{---}          & too massive    & \multirow{2}{*}{too massive/young} & mass too small\\
the captured star?      & \multicolumn{1}{c}{}   & for its SMA           & \multicolumn{1}{c}{} & for its SMA\\
\hline
\end{tabular}
\end{table*}
\renewcommand{\arraystretch}{1}

\section{summary}\label{sec:summary}
Due to its large ejection speed of 1800 km/s, the hyper-velocity star S5-HVS1, was most likely produced by tidal break-up of a tight binary system by SgrA* \citep{1988Natur.331..687H}. We use general arguments to constrain the properties of the captured companion of S5-HVS1. The mass of the bound star is in the range $(1.2,\, 6)\msun$. It was initially in a highly eccentric orbit with typical pericenter distance 1--10~AU and semimajor axis $\sim 10^3\rm \,AU$. We explore whether it survived until today by following its interior and orbital evolution for 5 Myr after the capture. 

For each pericenter passage, energy is injected into the star’s normal oscillation modes via tidal excitation according to linear perturbation. We assume that the mode energy (mainly in the g-modes and sometimes f-modes) is quickly dissipated into heat in the radiative zone and then study the response of the star's interior structure by $\mathtt{MESA}$ simulations. We find that cumulative tidal heating leads to runaway radial expansion and disruption if the pericenter distance is below a critical value of about 3 AU.

We then study the angular momentum diffusion in the dense stellar  environment near SgrA* using a simple Monte Carlo scattering simulation that captures the main physics. We find that the angular momentum of the captured star may evolve substantially from the initial value over the past 5 Myr, causing a large fraction of cases to be tidally disrupted and the pericenter distances of the surviving cases to be in the range 10 to 100 AU. Overall, we find the surviving probability to be between 5\% and 50\%, depending on the local relaxation time of the captured orbit and whether the loss cone is empty or full where the binary originated.

The ejection rate of S5-HVS1-like stars, somewhere between $10^{-7}$ and $10^{-5}\rm\,yr^{-1}$, is highly uncertain due to observational incompleteness and Poisson error. For a median rate of $\Gamma = 10^{-6}\rm\,yr^{-1}$, there are cumulatively between 0.25 and 2.5 surviving stars in similar orbits as the captured star studied in this work, which are potentially the most extreme members of the S-star cluster. 
% with pericenter distances between 10-100 AU, making them potentially the most extreme members of the S-star cluster.
They are detectable (typically $K_{\rm s}\lesssim 18.5\,$mag) by the GRAVITY instrument and by future Extremely Large Telescopes and will provide an extraordinary probe of the spin of SgrA*.

% The probability of the bound star surviving prompt or delayed tidal disruptions and being detectable by the GRAVITY instrument (at $K\lesssim 19\rm\,mag$) is $P_{\rm det}\sim 1/3$. The number of similarly captured stars accumulated in the past 100 Myr is estimated to be between $5$ to $500$. These stars are in highly relativistic orbits ($\rp \sim 100\rg$) and hence provide an exciting laboratory to measure the spin of the BH at GC (down to the precision of $\chi\sim 0.1$) and test general relativity. We also provide a lower limit on the main-sequence EMRI rate of $\sim 10^{-7}\rm\, yr^{-1}$, as contributed by the binary break-up channel only.

\section*{Data Availability}
The data underlying this article will be shared on reasonable request to the corresponding author.

\section*{acknowledgments}
We thank Reinhard Genzel, Aleksey Generozov, Cl{\'e}ment Bonnerot, Saul Teukolsky for useful discussions. WL was supported by the David and Ellen Lee Fellowship at Caltech. JF acknowledges support from an Innovator Grant from The Rose Hills Foundation, and the Sloan Foundation through grant FG2018-10515. HBP is greatful for the support from the Kingsley distinguished-visitor program at Caltech. TSL was supported by NASA through Hubble Fellowship grant HST-HF2-51439.001, awarded by the Space Telescope Science Institute, which is operated by the Association of Universities for Research in Astronomy, Inc., for NASA, under contract NAS5-26555. MWH acknowledges support by NASA through grant HST-GO-15199.001-A. TD was supported by NSF AAG AST-1909554.

{\small
\bibliographystyle{mnras}
\bibliography{refs}
}

\appendix
\section{Tidal coupling constant}\label{sec:tidal_coupling}
Consider a star of mass $M_*$ and radius $R_*$ in a parabolic orbit with pericenter distance $\rp = \chi R_*(M/M_*)^{1/3}$. The spatial coupling between a stellar oscillation mode $nl$ and the tidal potential is given by \citep{2012MNRAS.421..983B, 2017MNRAS.472.1538F}
\begin{equation}
    Q_{nl} = - (2l + 1) \delta \Phi_{nl}(R_*)/(4\pi),
\end{equation}
where $\delta\Phi_{nl}(R_*)$ is the Eulerian gravitational perturbation at the stellar surface in units of $GM_*/R_*$, and $\omega_n$ is the frequency of the eigenmode in units of $(R_*^3/GM_*)^{1/2}$. For each $\mathtt{MESA}$ stellar model, we compute $\delta\Phi_{nl}(R_*)$ and hence $Q_{nl}$ with $\mathtt{GYRE}$ \citep{2013MNRAS.435.3406T}.
The temporal coupling between the orbit and the stellar oscillation is given by
\begin{equation}
    K_{nlm} = {W_{lm}\over 2\pi} 2^{3/2} \chi^{3/2} I_{lm}(\chi^{3/2}\omega_n),
\end{equation}
where $\chi$ is the ratio between pericenter distance and tidal disruption radius, $W_{20} = -\sqrt{\pi/5}$, $W_{2\pm2} = \sqrt{3\pi/10}$, and the function $I_{lm}(y)$ is given by equation (43) of \citet{1977ApJ...213..183P}. Then the tidal coupling constant is given by
\begin{equation}
    T (\chi, M_*) = 2\pi^2\sum|Q_{nl}|^2 |K_{nlm}|^2.
\end{equation}
Our $\mathtt{MESA}$ simulations including tidal heating are based on cubic polynomial fits to $\mr{log}[T(\chi)]$ for each stellar mass considered, which are accurate to within 2\%.

\begin{figure}
  \centering
\includegraphics[width = 0.45 \textwidth]{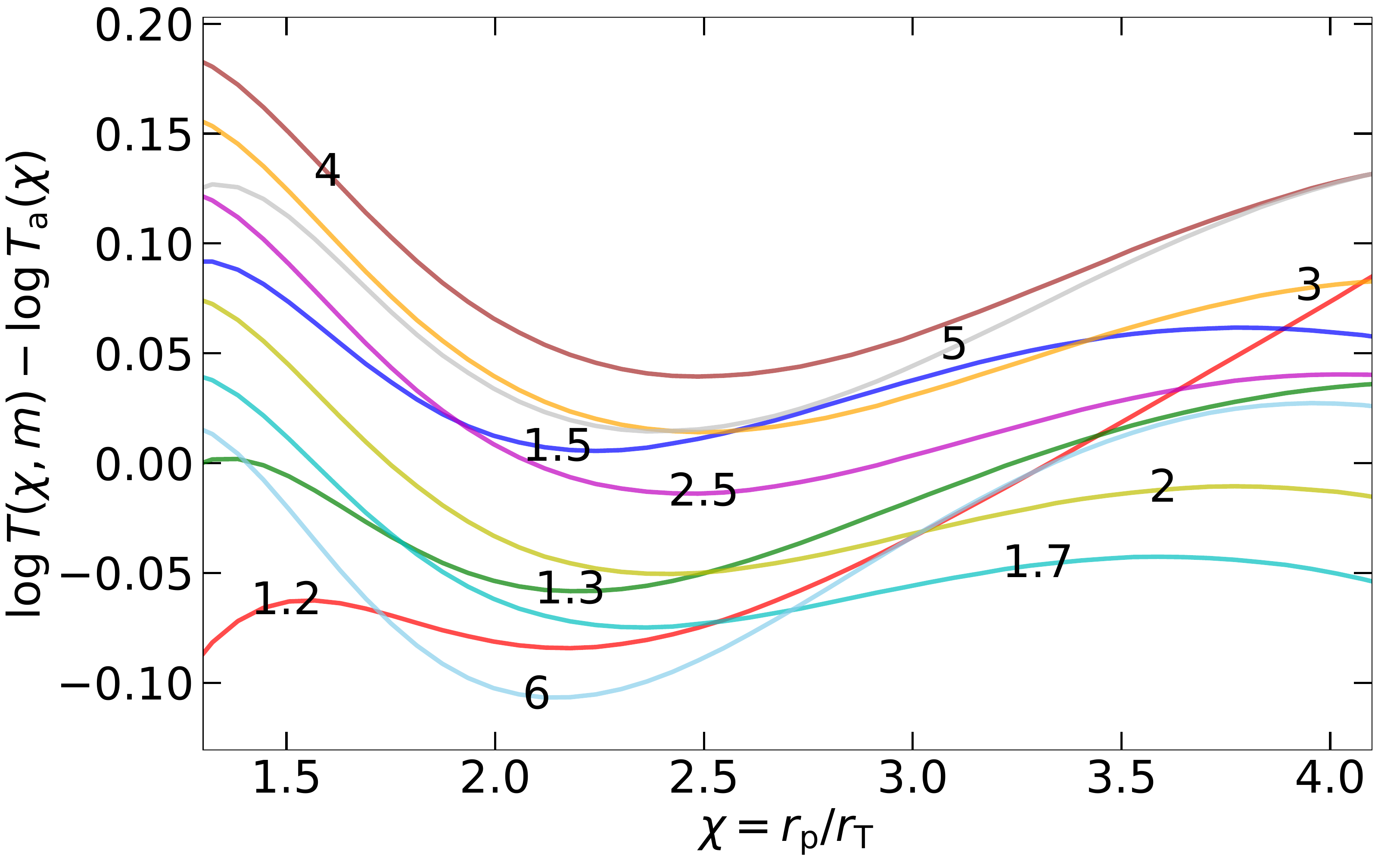}
\caption{Tidal coupling coefficient $T(\chi,M_*)$ for different stellar masses (in $\msun$) marked on each line. Our numerical $\mathtt{MESA}$ simulations are based on accurate $T(\chi,M_*)$ for each stellar mass.  Here, for intuitive understanding, we show the residual after subtracting the simple analytical expression of $\mr{log}\,T_{\rm a}(\chi)$ in eq. (\ref{eq:T_chi_analytic}), which provides a good description to within $15\%$ for for $1.5<M_*<6\msun$ and $1.5 < \chi < 4$. The coupling coefficients for lower masses $M_*< 1.5\,\msun$ are larger due to their more convective (and less centrally concentrated) structures, so we have added $\Delta \mr{log}\, T_{\rm a}=0.2$ (or 0.55) to eq. (\ref{eq:T_chi_analytic}) for the case of $M_*=1.3\msun$ (or $1.2\msun$). The stellar models all have main-sequence age of $50\,$Myr and $Z=0.03$.
}\label{fig:T_chi}
\end{figure}

For intuitive understanding, we find that the tidal coupling constant can be roughly described by the following simple analytical expression
% The numerical function can be roughly described by the following analytical form
\begin{equation}\label{eq:T_chi_analytic}
    % T(\chi, M_*) \approx 0.09\,\mr{exp}(-1.95\chi)
    \mr{log}\,T(\chi, M_*) \simeq -0.9(\chi+1),
\end{equation}
valid to within 15\% for $1.5<M_*<6\msun$ and $1.5 < \chi < 4$, as shown in Fig. \ref{fig:T_chi}. Therefore, the dimensionless tidal energy deposition per pericenter passage is
\begin{equation}
    \Delta \t{E} = {\Delta E\over GM_*^2/R_*} \simeq \chi^{-6} 10^{-0.9(\chi+1)},
\end{equation}
which means $\Delta \t{E}(\chi=1.5)\approx 4.9\times10^{-4}$, $\Delta \t{E}(2)\approx 3.1\times10^{-5}$, $\Delta \t{E}(3)\approx 3.4\times10^{-7}$, and $\Delta \t{E}(4)\approx 7.7\times10^{-9}$.

\label{lastpage}
\end{document}